\documentstyle[12pt,fullpage,epsf]{article}
\begin{document}

\def\balpha{\mbox{\boldmath$\alpha$}}
\def\bgamma{\hbox{\twelvembf\char\number 13}}
\def\bsigma{\hbox{\twelvembf\char\number 27}}
\def\bepsilon{\hbox{\twelvembf\char\number 15}}
\def\sgone{{\cal S}_1{\cal G}}
\def\sgtwo{{\cal S}_2{\cal G}}
\def\ob{\overline}
\def\bx{{\bf x}}
\def\by{{\bf y}}
\def\bz{{\bf z}}
\def\D{{\cal D}}
\def\R{{\cal R}}
\def\Q{{\cal Q}}
\def\G{{\cal G}}
\def\O{{\cal O}}
\def\N{{\cal N}}
\def\P{{\cal P}}
\def\Nlarge{N_c\rightarrow\infty}
\def\Tr{{\rm Tr}}
\newcommand{\ket}[1]{|#1\rangle}
\newcommand{\bra}[1]{\langle#1|}
\newcommand{\firstket}[1]{|#1)}
\newcommand{\firstbra}[1]{(#1|}

\pagestyle{plain}

\hfill hep-th/9607183

\hfill July 1996

\vspace{1cm}
\begin{center}
{\LARGE\bf Bits of String and Bits of Branes}

\vspace{2cm}

Oren Bergman\footnote{E-mail  address: oren@phys.ufl.edu}
\vskip 0.5cm
{\it Institute for Fundamental Theory\\
Department of Physics, University of Florida\\
Gainesville, FL 32611}

\vspace{2cm}

{\it Presented as a poster at the Strings '96 conference

``Current Trends in String Theory''

Santa-Barbara, CA, July 1996}

\end{center}

\vspace{3cm}
\begin{center}
{\bf ABSTRACT}
\end{center}

\noindent String-bit models are both an efficient way of organizing
string perturbation theory, and a possible non-perturbative composite
description of string theory. 
This is a summary of ideas and results in string-bit
and superstring-bit models, as presented in the  Strings '96 
conference.

\vfill
\newpage

\Large

\centerline{\LARGE\bf What are string-bits?}
\vspace{0.5cm}
Technically speaking, string-bits are : [1--4]
\begin{itemize}
\item {point particles in $d$ space $+$ $1$ time dimensions, 
transforming in the
adjoint representation of $U(N_c)$.}
\item {subject to Galilean invariant (nonrelativistic) dynamics.}
\item {able to form closed chain bound states, which are $U(N_c)$ singlets.}
\end{itemize}

\begin{figure*}[htb]
\epsfxsize=4.5in
\centerline{\epsffile{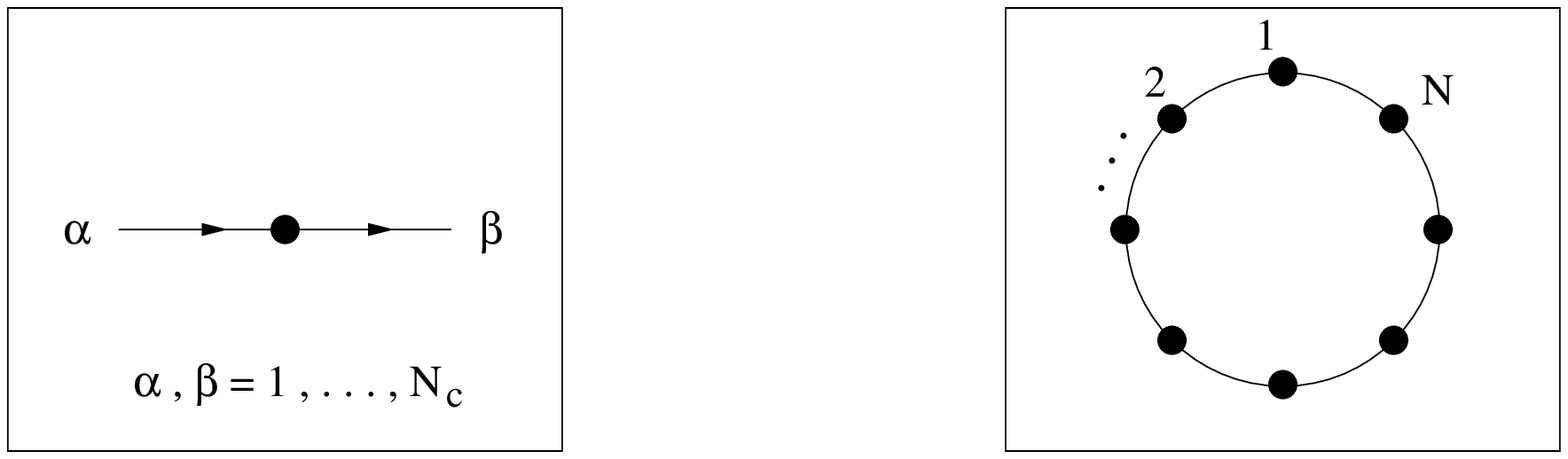}}
\end{figure*}

\noindent These properties guarantee that in the continuum limit :
$$N\rightarrow\infty\;\; , \;\; 
  mN={\rm constant}\; ,$$
\begin{itemize}
 \item {$N_c\rightarrow\infty$ corresponds to free relativistic 
        light-cone string theory in $D=d+2$ dimensions.}
 \item {The $1/N_c$ expansion corresponds to string perturbation 
        theory.}
\end{itemize}
\vspace{0.5cm}
\noindent In this context,
\begin{itemize}
\item {The $d$-dimensional Galilei group is understood as the
light-cone subgroup of the $d+2$-dimensional Poincar\'e group.}
\item {The string coupling constant is given by $1/N_c$.}
\end{itemize}

\newpage
\noindent Philosophically, there are two answers to this question, a
conservative one and a radical one :
\begin{enumerate}
 \item {String-bits are a useful way of thinking about perturbative
 (light-cone) string theory. String-bit models incorporate
 both free string dynamics and perturbative string interactions at
 large $N_c$, and
 therefore offer an extremely efficient discretization of string theory.
 Key features : }
 \begin{itemize}
  \item {Bosonic string-bit model with continuum properties of 
         bosonic string theory.}
  \item {SUSY string-bit model with continuum properties of 
      type IIB superstring theory.}
  \item {String interaction vertices, including necessary contact
      terms for the supersrtring, directly from string-bit interactions.}
  \item {Logarithmic growth of free strings with energy.}
 \end{itemize}
 \item {String-bits are the fundamental degrees of freedom of string theory.
 String-bit models describe the dynamics of string constituents for any
 value of $N_c$, and therefore represent possible non-perturbative
 formulations of string theory. Key features : }
 \begin{itemize}
  \item {String theory is a low energy effective theory.}
  \item {Dimensional reduction : $d+2 \rightarrow d+1$.}
  \item {Gauge (diffeomorphism) and Poincar\'e invariance are 
       abandoned in favor of a global $U(N_c)$ and Galilean
       invariance.}
  \item {Stability of discrete chains of bits from SUSY.}
  \item {Free energy and dissociation transition at high 
        temperature.}
  \item {The physical size of an interacting string?}
 \end{itemize}
\end{enumerate}

\newpage

\centerline{\LARGE\bf String-bit models and perturbative string theory}
\vspace{0.5cm}
\noindent\underline{\bf Bosonic model} [1,2]

\vspace{0.5cm}
\noindent The string-bit dynamics are described by a Galilean invariant
$N_c\times N_c$ matrix field theory with hamiltonian :
$$
H = {1\over 2m}\int dx{\rm Tr}|\nabla\phi|^2
  + {T_0^2\over 2mN_c}\int dxdy V(x-y){\rm Tr}[\phi^\dagger(x)
    \phi^\dagger(y)\phi(y)\phi(x)] 
$$
The matrix field $\phi(x)_\alpha^\beta$ acts as an annihilation
operator for a string-bit, and its conjugate 
$\phi^\dagger(x)_\alpha^\beta$ acts as a creation operator.
The mass of each bit is $m$.

\vspace{0.5cm}
\noindent A single closed chain of $N$ bits is described by the Fock state :
$$
\ket{\psi_N} = \int dx_1\cdots dx_N
             {\rm Tr}[\phi^\dagger(x_1)\cdots
             \phi^\dagger(x_N)]\ket{0}\psi_N(x_1,..,x_N) \; ,
$$
where the wavefunction $\psi_N(x_1,..,x_N)$ is cyclically symmetric.
The action of $H$ on this state produces two terms
corresponding to a single-chain state and a two-chain state :
$$
H\ket{\psi_N} = \ket{h\psi_N} 
              + {1\over N_c}\sum_{k=1}^N\ket{v_k\psi_{k,N-k}} \; ,
$$
\begin{figure*}[h]
\epsfxsize=5in
\centerline{\epsffile{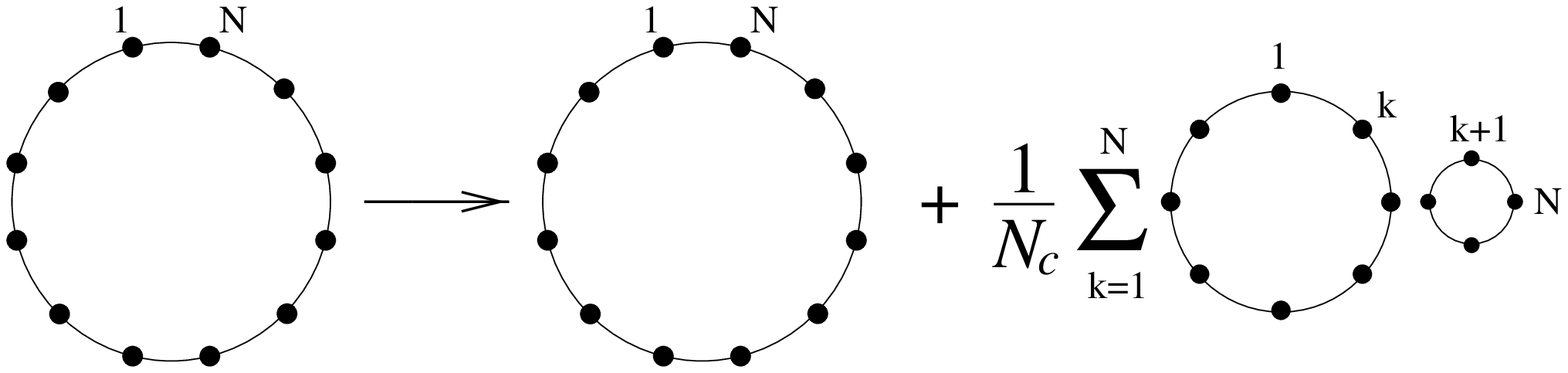}}
\end{figure*}

\noindent where $h$ is the coordinate space (first-quantized) hamiltonian 
$$
h = {1\over 2m}\sum_{k=1}^N
  \Big[p_k^2 + T_0^2V(x_{k+1}-x_k)\Big]\; ,
$$
and $v_k={T_0^2\over 2m}V(x_{k+1}-x_1)$.

\newpage
\noindent\underline{$N_c\rightarrow\infty$} :

\noindent Single-chian $N$-bit wavefunction is an eigenstate of $h$,
$$
h\psi_N(x_1,\ldots,x_N) = E\psi_N(x_1,\ldots,x_N) \; ,
$$
$\Longrightarrow$ Physical bound chain $\Longleftrightarrow$
$V(x)$ strong enough to bind.

\vspace{0.5cm}
\noindent Harmonic model $V(x)=x^2$ exactly soluable, excitation spectrum :
$$
E_n = {2T_0\over m}\sin{n\pi\over N} \; .
$$
In the continuum limit the hamiltonian becomes :
$$
h \rightarrow {1\over 2T_0}\int_0^{mN/T_0}
 d\sigma \Big[{\cal P}(\sigma)^2 
 + T_0^2 x^\prime(\sigma)^2\Big] \; .
$$
Finite energy modes occur for $n\ll N$ and for $N-n\ll N$ :
$$
E_n \rightarrow {2\pi nT_0\over mN}\;\; , \;\; 
E_n \rightarrow {2\pi (N-n)T_0\over mN} \; .
$$
$\Longrightarrow$ Light-cone hamiltonian ($p^-$) and 
right and left-moving spectra of bosonic string, where
the longitudinal momentum is $p^+ = mN$.
The extra dimension $x^-$ then emerges as the conjugate of $p^+$. 

\vspace{0.5cm}
\noindent\underline{Large $N_c$} :

\noindent Two-chain term gives a 1-chain
$\leftrightarrow$ 2-chain transition at $\O(1/N_c)$ :
$$
\bra{\psi_N}H\ket{\psi_L,\psi_{N-L}} ~
  {\raisebox{1ex}{$\stackrel{\mbox{\tiny $N\rightarrow\infty$}}
    {\raisebox{-1ex}{$\sim$}}$}} ~
  {1\over N_c}(mT_0)^{-3/2}N^{3/2 - d/8} \; ,
$$
$\Longrightarrow$ finite in continuum $\Longleftrightarrow$ $d=24$
$\Longleftrightarrow$ $D=26$.
\begin{figure*}[htb]
\epsfxsize=2.8in
\centerline{\epsffile{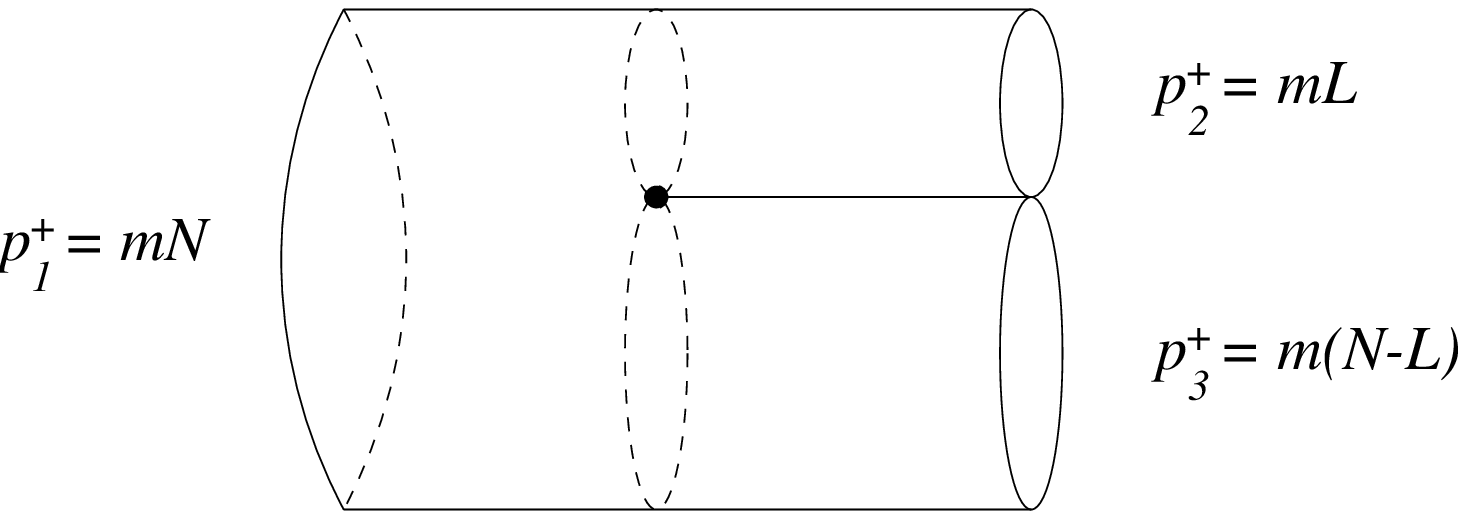}}
\end{figure*}

\noindent We stress that this vertex was derived from the same 
term in $H$ which
gave the {\bf free} string tension $T_0$.

\newpage
\noindent\underline{Limitations} :

\noindent This bosonic model is of little use however, because :
\begin{enumerate}
 \item Chains are unstable, and decay into smaller chains through
  the above vertex. 
  \begin{figure*}[htb]
  \epsfxsize=4in
  \centerline{\epsffile{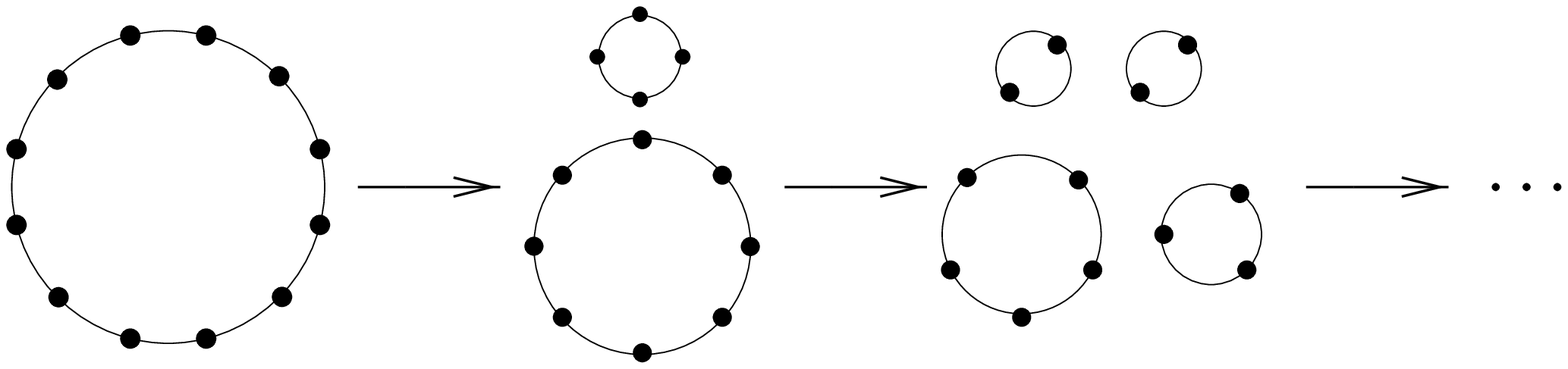}}
  \end{figure*}
  
  \noindent The ground state energy of a long chain is :
  $$
  E_{\rm G.S.} = d\sum_{n=1}^{N-1}E_n 
   = (d/m)\left[aN + {b\over N}
       + O({1\over N^2})\right] \; .
  $$
  In this model $b<0$, so two chains are lighter than one, hence
  the instability. Continuum interpretation : 
  $b = M^2_{\rm tachyon}$.
 \item Long range interaction between separate chains,
  \begin{figure*}[htb]
  \epsfxsize=4in
  \centerline{\epsffile{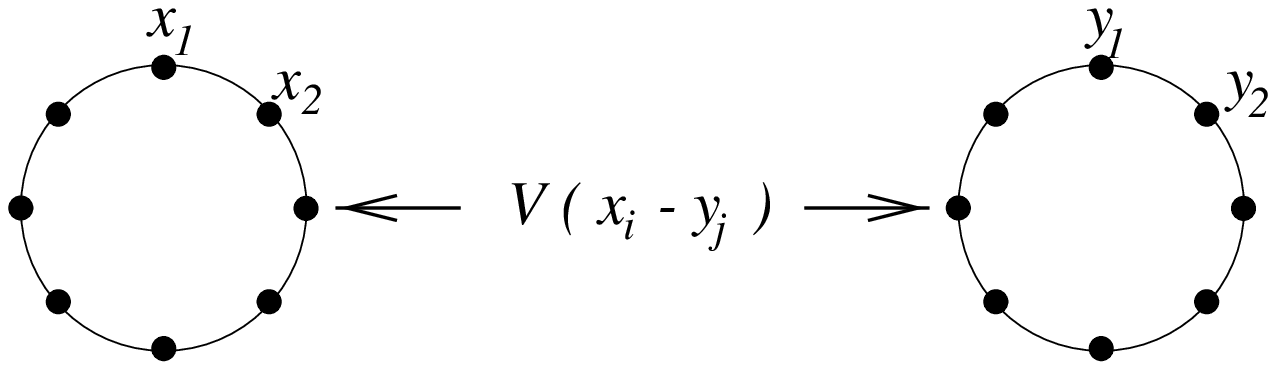}}
  \end{figure*}
  
  \noindent $V(x_i-y_j)\sim (x_i-y_j)^2/N_c^2$, precludes
  a well defined S-matrix.
 \item The bit interaction term in $H$ is not unique,
  e.g. add to it :
  $$
  H^\prime_I = {\lambda\over N_c}\int dxdy U(x-y)
  :\Tr[\phi^\dagger(x)\phi(x)\phi^\dagger(y)\phi(y)]: \; .
  $$
  Different matrix ordering in the trace implies
  $$
  H^\prime_I\ket{\psi_N} = {1\over N_c}\sum_k\ket{u_k\psi_{k,N-k}} \; ,
  $$
  no $\O(1)$ single chain term. Same free string limit, 
  slightly different string interactions.
\end{enumerate}

\newpage
\noindent\underline{\bf Supersymmetric models} [3,4]

\vspace{0.5cm}
\noindent In continuum string theory SUSY resolves the first issue by
removing the Tachyon. SUSY string bit models will resolve
the first and second issues, and will reduce somewhat the 
the third. 

\vspace{0.5cm}
\noindent First, we extend the Galilei group to an $\N=1$
Super-Galilei algebra by adding two supercharges $\Q,\R$,
transforming as spinors under $SO(d)$ and satisfying among
other things
$$
 \{\Q^A,\Q^B\} = mN\delta^{AB}, \;
 \{\Q^A,\R^{\dot B}\} = {1\over 2}\alpha_i^{A\dot B}P^i, \;
   \{\R^{\dot A},\R^{\dot B}\} = \delta^{{\dot A}{\dot B}}H/2.
$$
For simplicity, let $d=1$ $\Longrightarrow$ drop spinor indices.

\vspace{0.5cm}
\noindent The model is defined by the supercharges :
\begin{eqnarray}
 {{\cal Q}} \hspace{-5pt} &\sim& \hspace{-7pt} \int \hspace{-5pt}dx~
    {{\rm Tr}}\phi^\dagger(x)\psi(x) + {\rm h.c.}
    \nonumber\\
 {{\cal R}} \hspace{-5pt}  &\sim& \hspace{-7pt} \int \hspace{-5pt}dx~
    {\rm Tr}\phi^\dagger(x)\psi^\prime(x) + 
    {1\over N_c}\int \hspace{-5pt}dx dy ~W(y-x) 
    {\rm Tr}\phi^\dagger(x)\rho(y)\psi(x) 
     + {\rm h.c.}, \nonumber
\label{secondquant}
\end{eqnarray}
where $\psi$ is the fermionic matrix field, and 
$\rho = [\phi^\dagger\phi +\psi^\dagger\psi]$ is 
the bit density matrix.
The hamiltonian is computed from the superalgebra:
\begin{eqnarray}
H &=& {1\over 2m}\int dx\Tr\Big[|\nabla\phi|^2 
                          + |\nabla\psi|^2\Big] \nonumber\\
&+& {1\over 2mN_c}\int dxdy\biggl\{
  \Big[W^2(y-x)+W^\prime(y-x)\Big]\Tr\phi^\dagger(x)\rho(y)\phi(x)
  \nonumber\\
&&~~~~~~~~~~~~~~~~~~~+ {\mbox{ other two-body terms}}\biggr\}\nonumber\\
&+& {1\over 2mN_c^2}\int dxdydz \biggl\{W(y-x)W(z-x):\Tr
    \phi^\dagger(x)\rho(z)\rho(y)\phi(x): \nonumber\\
&&~~~~~~~~~~~~~~~~~~~+ {\mbox{ other three-body terms}}\biggr\}\; .\nonumber
\end{eqnarray}
Note the presence of three-body terms, which were absent in the bosonic
model, but are required by supersymmetry. These will give rise to
superstring ``contact terms'' in the continuum limit.

\newpage
\noindent The action of $H$ on a single SUSY chain state,
$$
\ket{\Psi_N} = \int \prod dx_id\theta_i ~{\rm Tr}
  \prod [\phi^\dagger(x_i) + \psi^\dagger(x_i)\theta_i]
         \ket{0}\Psi_N(x_1,\theta_1\ldots ,x_N,\theta_N) 
$$
gives rise to several single-chain states, two-chain states and three-chain
states. 

\vspace{0.5cm}
\noindent\underline{$N_c\rightarrow\infty$} :

\noindent Only the original
single chain survives, acted on by the super-coordinate space hamiltonian :
\begin{eqnarray}
 h &=& {1\over 2m}\sum_{k=1}^N\biggl\{
   p_k^2 + W^2(x_{k+1}-x_k) \nonumber\\
 +&&\hspace{-30pt} W^\prime(x_{k+1}-x_k)\big[\theta_k\pi_{k}
                                 - \pi_k\theta_{k}
 +\pi_{k+1}\theta_k - \theta_{k+1}\pi_k-i(\theta_k\theta_{k+1}
+\pi_k\pi_{k+1})\big] \biggr\}. \nonumber
\end{eqnarray}

\vspace{0.5cm}
\noindent $W(x)=T_0x$ $\Longrightarrow$ SUSY harmonic model, exactly soluable.
\begin{itemize}
  \item ``statistics'' modes spectrum $=$ phonon spectrum.
  \item $E_{\rm G.S.}=0$ $\Longrightarrow$ chains are stable.
    We stress
    that this is more than required for stability of continuum strings,
    and holds for all $N$.
\end{itemize}

\vspace{0.5cm}
\noindent Changing to more familiar fermionic variables,
$$
S_k = {1\over\sqrt 2}(\theta_k + \pi_k) \;\; , \;\;
\tilde{S}_k  = {i\over\sqrt 2}(\theta_k - \pi_k) \; ,
$$
the continuum limit becomes :
$$
 h \rightarrow {1\over 2T_0}\int_0^{p^+/T_0}
 d\sigma \Big[{\cal P}(\sigma)^2 + T_0^2 x^\prime(\sigma)^2
 -iT_0S(\sigma)S^\prime(\sigma) 
 +iT_0\tilde{S}(\sigma)\tilde{S}^\prime(\sigma)\Big] \; ,
$$
$\Longrightarrow$ Light-cone hamiltonian of {\bf type IIB
superstring}.

\vspace{0.5cm}
\noindent Note that $\N=1$ SUSY $\rightarrow$ $\N=2$ SUSY in the 
continuum limit, since discretization breaks half the supersymmetry.

\newpage
\noindent\underline{Large $N_c$} :

\noindent As in the bosonic model, the non-nearest-neighbor
bit interactions will give rise to light-cone superstring vertices.
At $\O(1/N_c)$ we get a 3-string vertex :
\begin{figure*}[htb]
\epsfysize=1.3in
\centerline{\epsffile{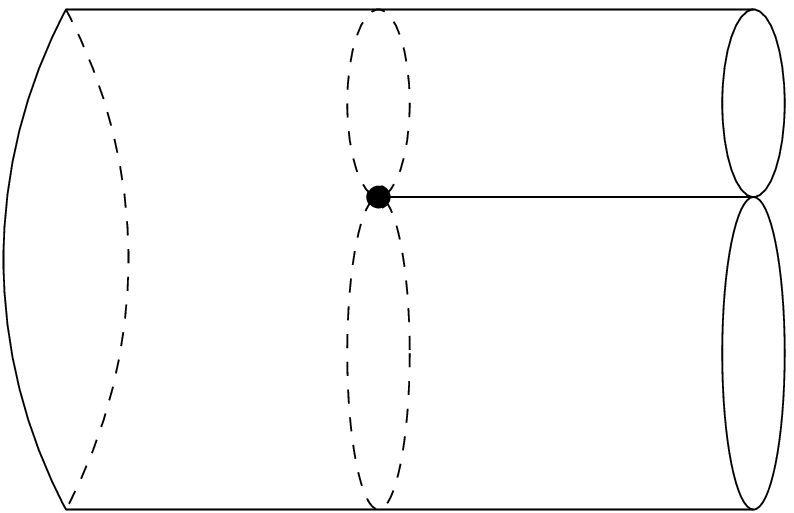}}
\end{figure*}

\noindent At $\O(1/N_c^2)$ the three body interactions will give rise to a 
4-string contact term
corresponding to the
boundary of the moduli space of the 4-string amplitude :
\begin{figure*}[htb]
\epsfysize=1.3in
\centerline{\epsffile{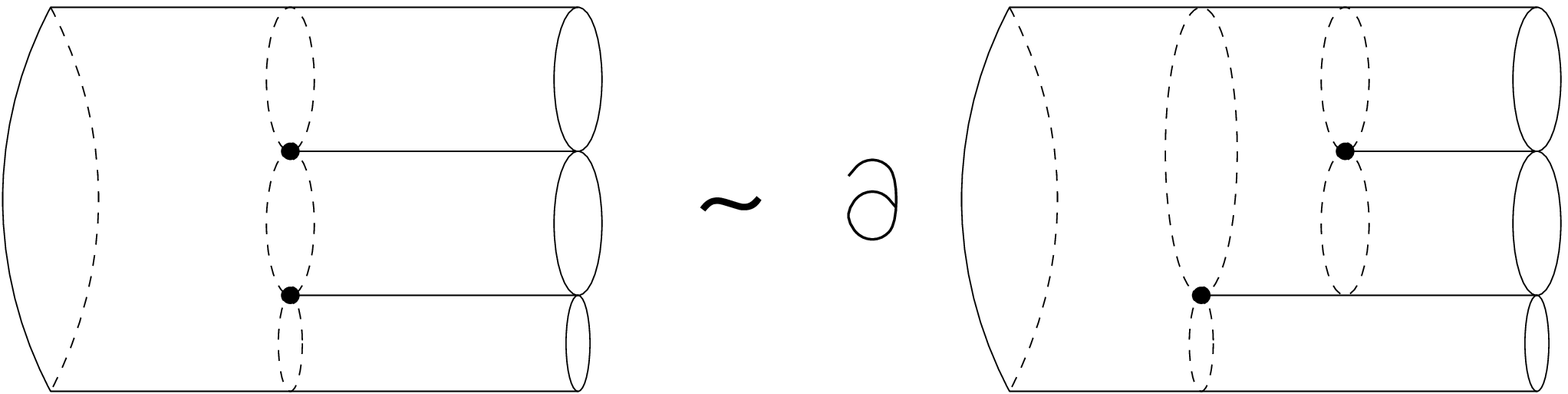}}
\end{figure*}

\noindent and a 2-string
contact term corresponding to the boundary of the moduli space of the 
one loop string propagator : 
\begin{figure*}[htb]
\epsfysize=1.3in
\centerline{\epsffile{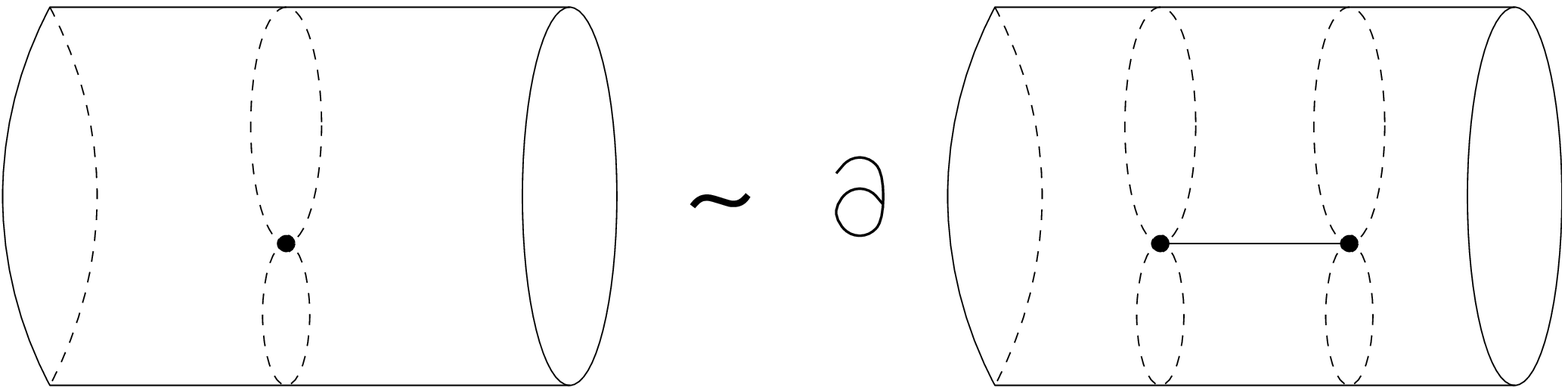}}
\end{figure*}

\noindent Such contact terms (at least the 4-string vertex) were shown to
be necessary in interacting superstring theory by supersymmetry [5]. 
Here they arise naturally, just like
the usual 3-string vertex, from the superstring-bit
model.

\newpage
\noindent\underline{Short range interactions} :

\noindent SUSY harmonic model still suffers from long range interactions
between separate chains. Can we make the interaction short range, while still
maintaining a stringy continuum limit?

\vspace{0.5cm}
\noindent Yes! In the $d=1$ superstring-bit model we can prove a 
restricted form of UNIVERSALITY. Let :
$$
W(x) = T_0x + \delta W(x)\; .
$$
As long as $|\delta W(x)| \ll T_0 |x| \;\; {\rm for}\; 
|x|<\sqrt{\alpha^\prime}$, the only effect is to renormalize
the string tension :
$$
T_0\rightarrow T_0
   + \langle\delta W^\prime(x_2-x_1)\rangle \; .
$$
So we deform the superpotential as in the figure,
\begin{figure*}[htb]
\epsfxsize=1.8in
\centerline{\epsffile{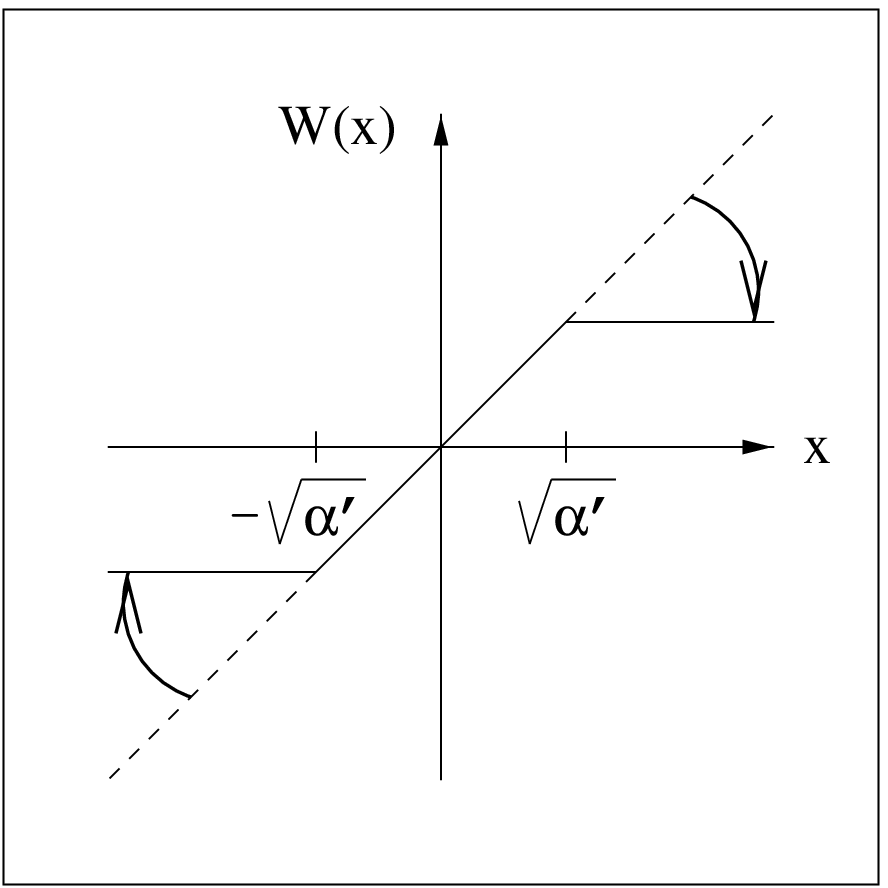}}
\end{figure*}
resulting in a short range interaction. This is not quite enough
for a well defined S-matrix, since the chain separation energy
is still large (diverges in the continuum limit). We fix this
by replacing the density matrix :
$$
\rho(y) \rightarrow 
  \rho(y)-:[\phi\phi^\dagger-\psi\psi^\dagger]:
$$
in the equation for the supercharge $\R$. This has the effect
of freeing the two chains when they are far apart.

\vspace{0.5cm}
\noindent The requirement of asymptotic freedom also has the effect 
of reducing the non-uniqueness of the bit interactions, since only very
special interactions will have this property.

\newpage
\centerline{\LARGE\bf String-bits as the fundamental degrees of freedom}
\begin{itemize}
 \item Continuum limit equivalent to low energy limit, so string
  theory can be thought of as a low energy effective theory of a
  certain string-bit model.
 \item Stability at the discrete level allows such an interpretation.
 \item String bit model can be analyzed for small $N_c$,
  with possible implications on non-perturbative string behavior.
\end{itemize}

\noindent\underline{High temperature} : [6]

\noindent The binding energy of two bits in a chain is 
$E_B \sim T_0/m$, 
so a dissociation phase transition is expected to occur at 
$T_c\sim T_0/m$. The low temperature phase is a bound chain,
whereas the high temperature phase is a gas of nonrelativistic
weakly interacting particles in $d$ space dimensions, with a free
energy :
$$
{F\over VT} \sim T^{d/2} \; .
$$
For $d=2$, corresponding to a low temperature phase of 4-dimensional
string theory, this gives the result found by Atick and Witten [7]
using string perturbation theory. The linear dependence on temperature
implies that string theory ultimately contains far fewer degrees
of freedom than any relativistic field theory, even though it
is not manifest in the perturbative approach.
String-bits offer a possible physical realization of this idea.
\begin{figure*}[htb]
\epsfxsize=4.6in
\centerline{\epsffile{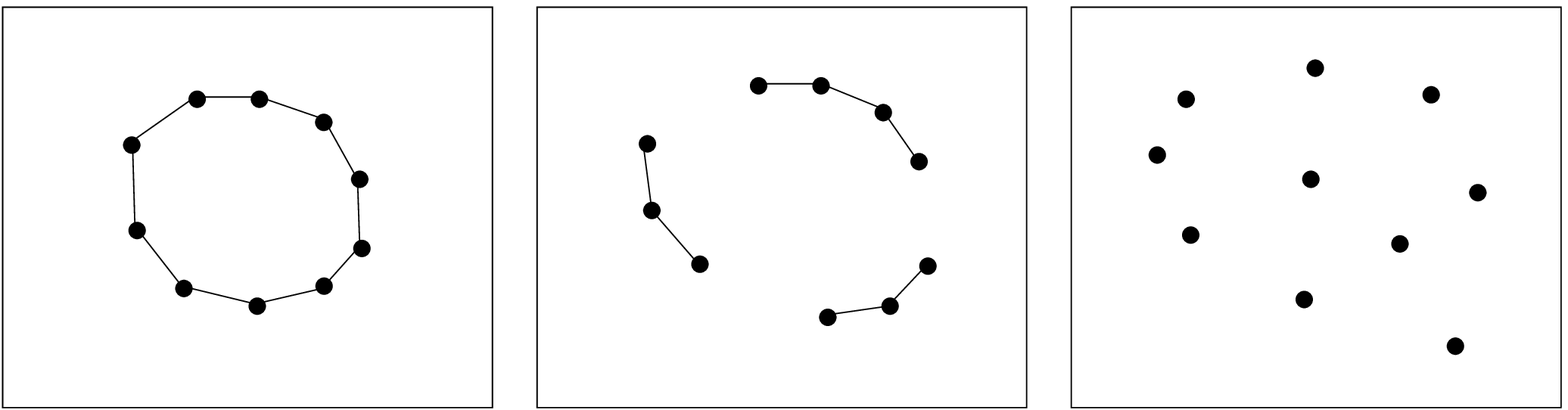}}
\end{figure*}

\newpage

\noindent\underline{String Growth} : [8]
\begin{itemize}
 \item Physical (transverse) size of perturbative string is infinite :
 $$
 R_\perp^2 \sim \sum_{n=1}^\infty {1\over n} \; .
 $$
 \item Finite resolution time $\epsilon$ $\Longrightarrow$
 $$
 R_\perp^2\sim\ln{p^+\over\epsilon} \; .
 $$
 String grows denser with increasing $p^+$, invalidating perturbation
 theory at the Planck density $T_0/g^2$.
 \begin{figure*}[htb]
 \epsfxsize=4.7in
 \centerline{\epsffile{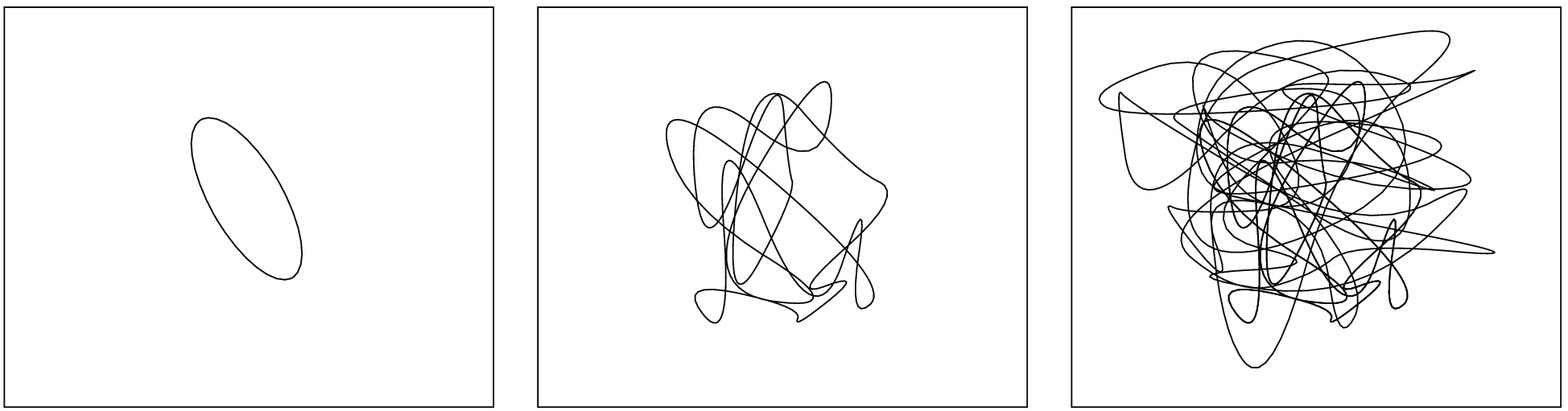}}
 \end{figure*}
 \item Black hole entropy $\propto A/\hbar$ $\Longrightarrow$
 seems to require $R_\perp^2\sim p^+/\epsilon$ at high $p^+$ [9].
 This would be a non-perturbative effect.
\end{itemize}
In the string-bit picture a measure for the size is :
$$
 R^2(N) = {1\over N}\sum_{k=1}^N\bra{\rm G.S.}(x_k-x_1)^2\ket{\rm G.S.} \; .
$$
\begin{itemize}
 \item $N_c\rightarrow\infty$ $\Longrightarrow$ In the harmonic model
 $R^2(N)\sim\ln N$, but this was shown (numerically) to hold for other
 nearest-neighbor interactions as well, in particular short-range ones.
 \item Finite $N_c$ : String-bit models imply non-nearest-neighbor
 repulsions, ``bits with elbows''. We studied a toy model for elbows,
 that uses harmonic nearest-neighbor attraction with $\delta$-function
 repulsions :
$$
 h = {1\over 2m}\sum_{k=1}^N
  \bigg[p_k^2 + T_0^2(x_{k+1}-x_k)^2 + 
  g^2\sum_{l\neq k}^N\delta(x_k-x_l)\bigg] \; .
$$
\end{itemize}

\newpage
\noindent We used a variational approach using wavefunctions that
are exact solutions to an harmonic problem. The variational 
parameters can be taken to be the mode frequencies $\omega_n$.
The variational energy and size are then :
\begin{eqnarray*}
 E &=& {1\over 2m}\left[
     \sum_{n}\omega_n + 
     \sum_{n}{\sin^2(\pi n/N)\over \omega_n} +
     {1\over 2}g^2N^2\sum_{k}
     \left(\sum_{n}{\sin^2(\pi nk/N)\over \omega_n}\right)^{-1}
     \right] \; , \\
 R^2 &=& {1\over N}\sum_{n=1}^{N-1}{1/\omega_n} \; . 
\end{eqnarray*}
Numerical solution of variational problem resulted in the following growth
patterns :
\begin{figure*}[htb]
 \epsfxsize=6.5in
 \centerline{\epsffile{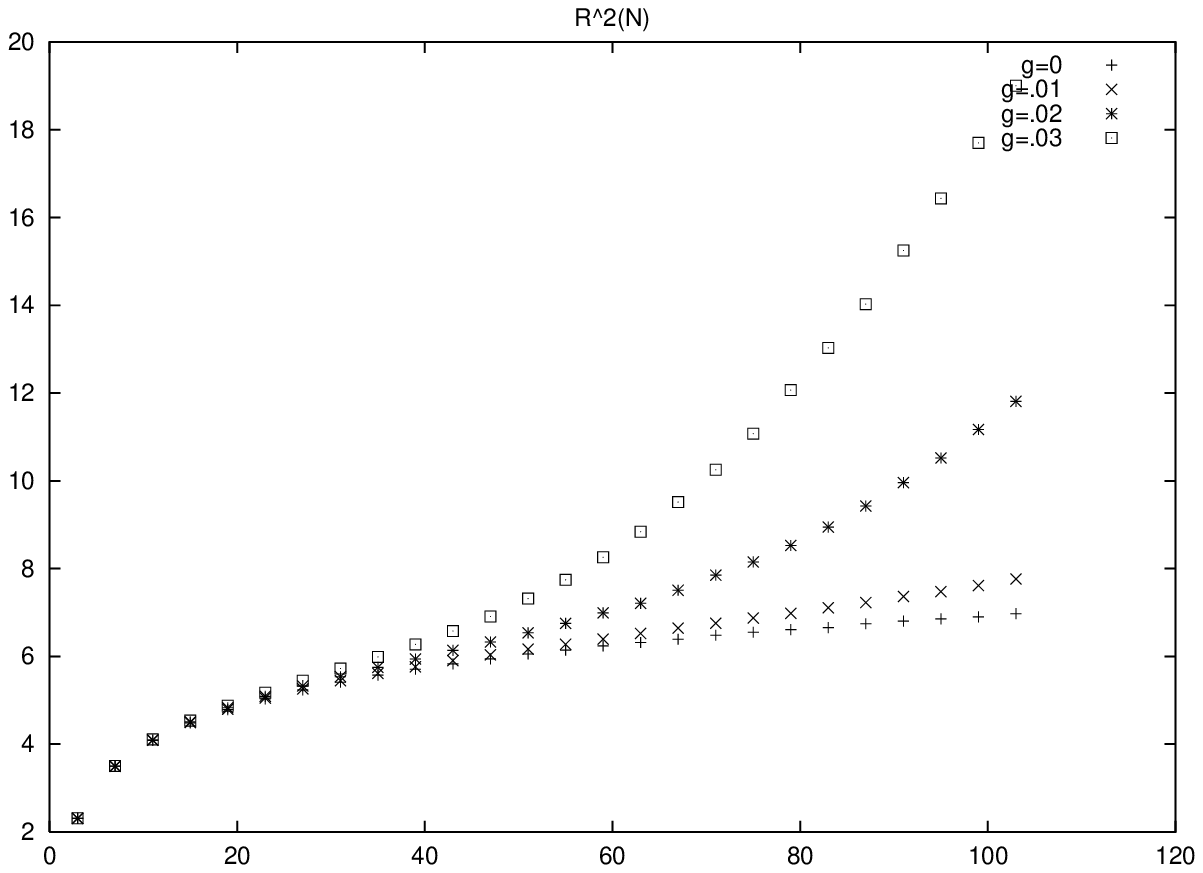}}
\end{figure*}

\noindent $\Longrightarrow$ Chains experience a small $N$ 
growth of $R^2\sim\ln N$, and
a large $N$ growth of $R^2\sim N^2$, {\bf not} the conjectured
$R^2\sim N\propto p^+$.

\noindent Somewhat discouraging, since relativistic string
$\Longleftrightarrow$ 
$R^2{\raisebox{-0.5ex}{$\stackrel{\mbox{\large $<$}}{\sim}$}}N$.

\newpage
\centerline{{\LARGE\bf Bits of Branes} [6]}
\begin{itemize}
 \item Membrane-bit field : $\phi(x)_{\alpha a}^{\beta b}$, four
       legged object.
 \item $U(N_c)\times U(N_c)$ global symmetry.
 \item Building blocks :
       \begin{figure*}[htb]
         \epsfxsize=5.5in
         \centerline{\epsffile{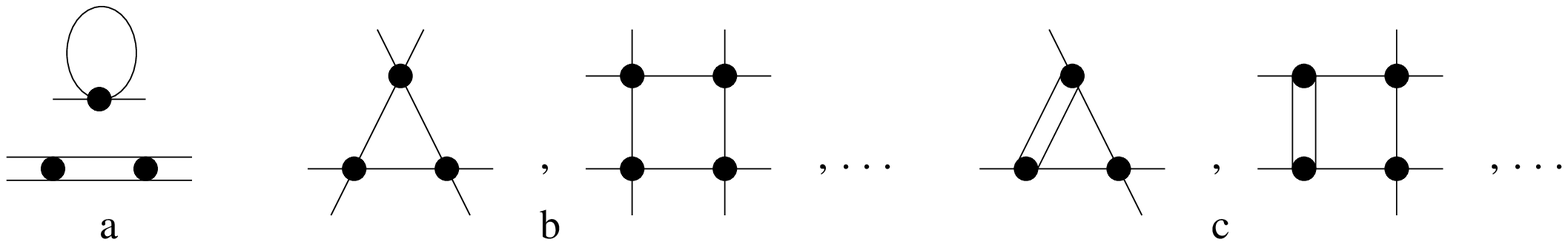}}
       \end{figure*}
       
   \noindent allow for a rich variety of singlet structures.
 \item Membranes made by tiling type (b) strucutres , e.g.
   \begin{figure*}[htb]
         \epsfxsize=5.5in
         \centerline{\epsffile{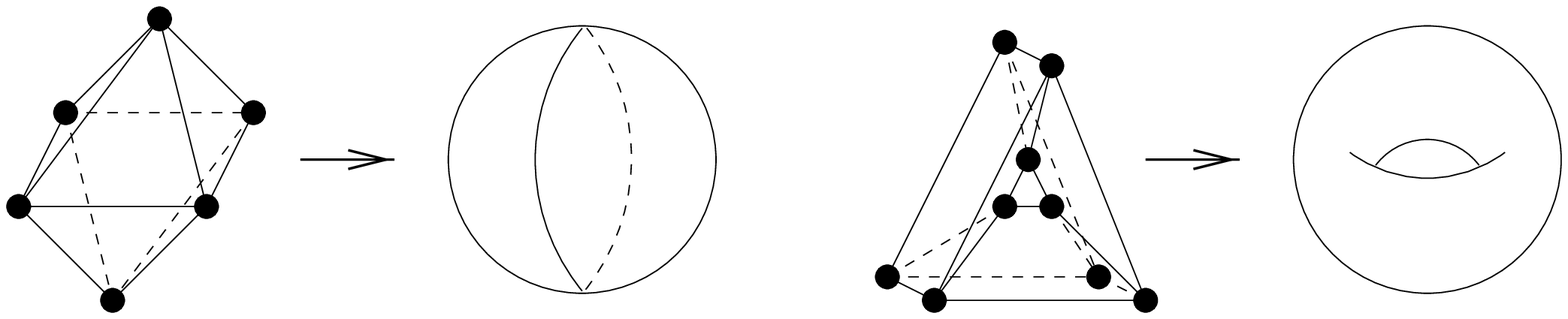}}
   \end{figure*}

 \item  Stringy objects made by connecting type (a) structures, e.g.
   \begin{figure*}[htb]
         \epsfysize=2in
         \centerline{\epsffile{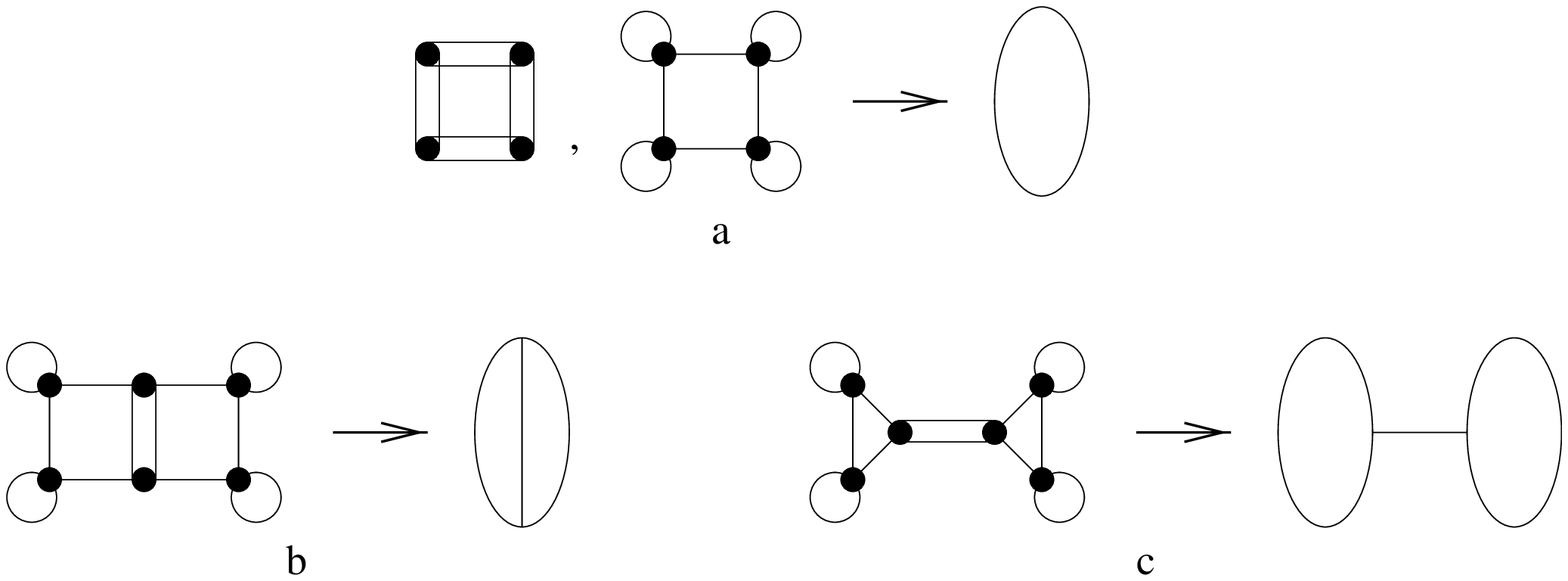}}
   \end{figure*}
 
 \noindent (a) Closed string.

 \noindent (b) Open string attached to closed string, 
               D-1-brane?

 \noindent (c) Open string attached to two closed strings,
               two D-1-branes?
\end{itemize}

\newpage
\noindent\underline{Dynamics} :
\begin{itemize}
 \item Free hamiltonian is unique :
      $$H_0 = {1\over 2m}\int dx 
      \nabla\phi^\dagger(x)_{\alpha a}^{\beta b}\cdot
      \nabla\phi(x)^{\alpha a}_{\beta b}\; .$$
 \item Unlike string-bit models, there are many possible membrane-bit
       interactions that yield a nearest-neighbor
       interaction pattern for $N_c\rightarrow\infty$, e.g.
  \begin{eqnarray*}
   H_1 &=& {1\over N_c^2}\int dxdy V_1(x-y)
   \phi^\dagger(x)_{\alpha a}^{\beta b}
   \phi^\dagger(y)_{\beta b}^{\gamma c}
   \phi(y)_{\gamma c}^{\delta d}
   \phi(x)_{\delta d}^{\alpha a} \\
   H_2 &=& {1\over N_c}\int dxdy V_2(x-y)
   \phi^\dagger(x)_{\alpha a}^{\beta b}
   \phi^\dagger(y)_{\beta c}^{\delta d}
   \phi(y)_{\delta d}^{\epsilon c}
   \phi(x)_{\epsilon b}^{\alpha a} \\
   H_3 &=& {1\over N_c}\int dxdy V_3(x-y)
   \phi^\dagger(x)_{\alpha a}^{\beta b}
   \phi^\dagger(y)_{\gamma b}^{\delta d}
   \phi(y)_{\delta d}^{\gamma c}
   \phi(x)_{\beta c}^{\alpha a} \; .
  \end{eqnarray*}
  Different interactions will generally give an $\O(1)$
  result for different kinds of singlet structures.
 \item Consider a membrane bit model defined by :
  $$
   H=H_0 + \lambda_1H_1 + \lambda_2H_2 + \lambda_3H_3 \; .
  $$
  It can be shown that if $\lambda_2=\lambda_3=0$, this
  model does not support pure membrane formation. At generic
  values of the parameters, both membranes and strings, as
  well as mixed structures are supported.

  \noindent $\Longrightarrow$ parameter space $\sim$ ``moduli'' space.
 \item $p$-brane-bits : $2p$-legged objects, can form $p$-branes,
       $p-1$-branes, etc.

   \noindent $\Longrightarrow$ Unified and {\bf democratic} description of 
             all $p$-branes as composites of bits.

\end{itemize}

\end{document}